\documentclass[journal]{IEEEtran}
\usepackage{cite}
\usepackage[pdftex]{graphicx}
\usepackage{amsmath}
\usepackage{bm}
\usepackage{amsfonts}

\usepackage{booktabs}
\usepackage{multirow}
\usepackage{float}
\usepackage{color}

\begin{document}
%
\title{CSformer: Bridging Convolution and Transformer for Compressive Sensing}

\author{Dongjie Ye,~\IEEEmembership{Graduate Student Member,~IEEE}, Zhangkai Ni,~\IEEEmembership{Graduate Student Member,~IEEE}, 
Hanli~Wang,~\IEEEmembership{Senior Member,~IEEE,} 
Jian~Zhang,~\IEEEmembership{Member,~IEEE,} 
Shiqi Wang,~\IEEEmembership{Senior Member,~IEEE}, 
and Sam Kwong,~\IEEEmembership{Fellow,~IEEE}
\thanks{Dongjie Ye, Zhangkai Ni, and Shiqi Wang are with the Department of Computer Science, City University of Hong Kong, Hong Kong 999077 (e-mail:dj.ye@my.cityu.edu.hk; eezkni@gmail.com; shiqwang@cityu.edu.hk).}
\thanks{Hanli Wang is with the Department of Computer Science \& Technology, Key Laboratory of Embedded System and Service Computing (Ministry of Education), and Shanghai Institute of Intelligent Science and Technology, Tongji University, Shanghai 200092, P. R. China (e-mail: hanliwang@tongji.edu.cn).}
\thanks{Jian Zhang is with the School of Electronic and Computer Engineering,
Peking University Shenzhen Graduate School, Shenzhen 518055, China, and
also with the Peng Cheng Laboratory, Shenzhen 518052, China (e-mail:
zhangjian.sz@pku.edu.cn).
}
\thanks{Sam Kwong is with the Department of Computer Science, City University of Hong Kong, Hong Kong 999077, and also with the City University of Hong Kong Shenzhen Research Institute, Shenzhen 518057, China (e-mail: cssamk@cityu.edu.hk).}
}

\markboth{Journal of \LaTeX\ Class Files,~Vol.~14, No.~8, August~2015}%
{Shell \MakeLowercase{\textit{\textit{et al.}}}: Bare Demo of IEEEtran.cls for IEEE Journals}


\maketitle

\begin{abstract}
Convolution neural networks (CNNs) have succeeded in compressive image sensing. However, due to the inductive bias of locality and weight sharing, the convolution operations demonstrate the intrinsic limitations in modeling the long-range dependency. Transformer, designed initially as a sequence-to-sequence model, excels at capturing global contexts due to the self-attention-based architectures even though it may be equipped with limited localization abilities. This paper proposes CSformer, a hybrid framework that integrates the advantages of leveraging both detailed spatial information from CNN and the global context provided by transformer for enhanced representation learning.  The proposed approach is an end-to-end compressive image sensing method, composed of adaptive sampling and recovery. In the sampling module, images are measured block-by-block by the learned sampling matrix. In the reconstruction stage, the measurement is projected into dual stems. One is the CNN stem for modeling the neighborhood relationships by convolution, and the other is the transformer stem for adopting global self-attention mechanism. The dual branches structure is concurrent, and the local features and global representations are fused under different resolutions to maximize the complementary of features. Furthermore, we explore a progressive strategy and window-based transformer block to reduce the parameter and computational complexity. The experimental results demonstrate the effectiveness of the dedicated transformer-based architecture for compressive sensing, which achieves superior performance compared to state-of-the-art methods on different datasets.

\end{abstract}


%
\IEEEpeerreviewmaketitle

\section{Introduction}
%
%
%
%
\IEEEPARstart{C}{ompressive} sensing (CS) theory demonstrates that a signal can be recovered from a much fewer acquired measurement than prescribed by Nyquist theorem with a high probability when the signal is sparse in certain transform domains \cite{donoho2006compressed}. The benefits of reducing sampling rate allow low-cost and efficient data compression, thereby relieving data storage and transmission bandwidth burden. These inherent merits enable it to be very desirable in a series of applications, including single-pixel camera \cite{duarte2008single}, magnetic resonance imaging \cite{lustig2007sparse,yang2017dagan}, video CS \cite{li2017structured}, and snapshot compressive imaging\cite{Yuan_2020_CVPR}.

In a compressive image sensing method, for the image $\mathbf{x} \in {\rm \mathbb{R}}^N$, the sampling stage first performs fast sampling of $\mathbf{x}$ to obtain the linear random measurements $\mathbf{y}=\mathbf{\Phi} \mathbf{x}\in {\rm \mathbb{R}}^M$. Here, $\mathbf{\Phi}\in {\rm \mathbb{R}}^{M\times N}$ is the sensing matrix with $M\ll N$, and $\frac{M}{N}$ denotes the CS sampling ratio. In the recovery stage, our goal is to infer the original image $\mathbf{x}$ given $\mathbf{y}$. Such inverse problem is typically under-determined because the number of unknowns $N$ is much larger than the number of observations $M$. To address this problem, traditional CS methods \cite{zhang2012image,zhang2014image,ahsen2017error} explore the sparsity as an image prior and find the sparsest signal among all measurements $\mathbf{y}$ by iteratively optimizing the sparsity-regularized problem. Although these methods usually have theoretical guarantees and simultaneously inherit interpretability, they inevitably suffer from the high computational cost dictated by the interactive calculations.

Compared to the conventional CS methods, neural networks have been leveraged to solve the image CS reconstruction problems by directly learning the inverse mapping from the compressive measurements to the original images. Recently, with the advent of deep learning (DL), diverse data-driven deep neural network models for CS have been shown to achieve impressive reconstruction quality and efficient recovery speed \cite{kulkarni2016,metzler2017learned,shi2017deep,zhang2018ista,yang2018admm,shi2019image,zhang2020optimization,zhang2020amp,you2021coast,you2021ista,sun2020learning}. In addition, the DL based CS methods often jointly learn the sampling and the reconstruction network to further improve the performance \cite{shi2017deep,you2021ista,zhang2020optimization,zhang2020amp}. 

In the existing CS literature, the DL based CS methods can be divided into two categories. The first is deep unfolding methods \cite{metzler2017learned,zhang2018ista,yang2018admm,zhang2020optimization,zhang2020amp,you2021ista}, which leverage the deep neural network to mimic the iterative restoration algorithms. They attempt to maintain the merits of both iterative recovery methods and the data-driven network methods by mapping each iteration into a network layer. The deep unfolding approaches can extend the representation capacity over iterative algorithms and avoid the limited interpretability of deep neural networks. However, since these methods are inspired by the traditional optimization processes, it inevitably limits the full potential of deep neural networks. 

The second group is the straightforward methods \cite{kulkarni2016,shi2017deep,xu2018lapran,shi2019image,sun2020dual,yao2019dr2} that are free from any handcrafted constraint. These methods can reconstruct images by one pass feed-forward of the learned convolutional neural network (CNN) given the measurement $\mathbf{y}$.  However, the principle of local processing limits CNN in trems of receptive fields and brings challenges in capturing long-range dependencies.  Moreover, the weight sharing of the convolution layer leads the interactions between images and filters to be content-independent \cite{liang2021swinir}. Numerous efforts have been devoted to addressing these problems, such as enlarging the kernel size of convolution, using multi-scale reconstruction, dynamic convolution, and the attention mechanism. Sun \textit{et al.} \cite{sun2020learning} explore the non-local prior to guide the network in view of the long-range dependencies problem. Furthermore, Sun \textit{et al.} \cite{sun2020dual} attempt to adopt dual-path attention network for CS, where the recovery structure is divided into structure and texture paths. Despite amplifying the ability of context modeling to some extent, these approaches are still unable to escape from the limitation of the locality, stranded by the CNN architecture.

Unlike prior convolution-based deep neural networks, transformer \cite{vaswani2017attention}, designed initially for sequence-to-sequence prediction in NLP domain, is well-suited to modeling global contexts due to the self-attention-based architectures. Inspired by the significant revolution of transformer in NLP, several researchers recently attempt to integrate the transformer into computer vision tasks, including image classification \cite{dosovitskiy2021an}, image processing \cite{chen2021pre,liang2021swinir,wang2021uformer}, and image generation \cite{jiang2021transgan}. With the simple and general-purpose neural architecture, transformer has been considered as an alternative to CNN and strived for better performance. However, a naïve application of transformer to CS reconstruction may not produce sufficiently competitive results that match the performance of CNN. The reason is that transformer can capture high-level semantics due to the global self-attention, which is helpful for image classification but lacks the low-level details for image restoration. In general, CNN has better generalization ability and faster convergence speed with its strong biases towards feature locality and spatial invariance, making it very efficient for the image. Transformer has higher model capacity thanks to less restriction by inductive biases, enabling self-attention layers to learn the inherent characteristics of larger datasets well. Moreover, the explosive computational complexity and colossal memory explosion for high resolution reconstruction are other challenges in applying transformer to CS.

To cope with the above issues and further refine the reconstruction quality, we propose CSformer, an effective and efficient transformer based method for image CS. CSformer integrates the advantages of leveraging both detailed spatial information from CNN and the global context provided by transformer. We design a hybrid framework that gradually increases the feature map resolution while reducing the dimension, enhancing the feature representation by multi-scale features while reducing memory cost and computational complexity. The proposed approach is an end-to-end compressive image sensing method composed of adaptive sampling and recovery. In the sampling module, images are measured block-by-block by the learned sampling matrix. In the reconstruction stage, we employ a progressive reconstruction strategy, and the CNN features are aligned with the layer-wise representations from the transformer. On one hand, the progressive reconstruction can process the multi-scale feature maps, which is helpful for representation learning and reduces the complexity of the parameters. On the other hand, CSformer enjoys the elaborate combination of local and global context by combining the two types of features at each resolution. Compared with the prevalent CNN-based methods, CSformer benefits from several aspects: (1) self-attention mechanism ensures the content-dependency between image and attention weight, (2) CNN provides a locality to transformer that lacks in addressing long-range dependencies, (3) progressive reconstruction balances the complexity and efficiency. To the best of our knowledge, CSformer is the first work to apply the transformer to CS. Experimental results demonstrate that our method has a promising performance and outperforms existing iterative methods and DL based methods.  

The main contributions of this work can be summarized as follows:
\begin{itemize}[\IEEEsetlabelwidth{Z}]
\item[1)] We propose CSformer, a hybrid framework that couples transformer with CNN for adaptive sampling and reconstruction of image CS. The proposed CSformer inherits both local features from CNN and global representations from transformer. 
\item[2)] To make full use of the complementary features of transformer and CNN, we introduce progressive reconstruction to aggregate the multi-scale features, which are thoughtfully designed for image CS to balance the complexity and performance with spatial variance.
\item[3)] Extensive experiments on various datasets demonstrate the superiority of the proposed CSformer. We reveal the great potential of transformer in combination with CNN for CS.
\end{itemize}

\begin{figure*}[htp]
\centering
\includegraphics[width=0.98\linewidth]{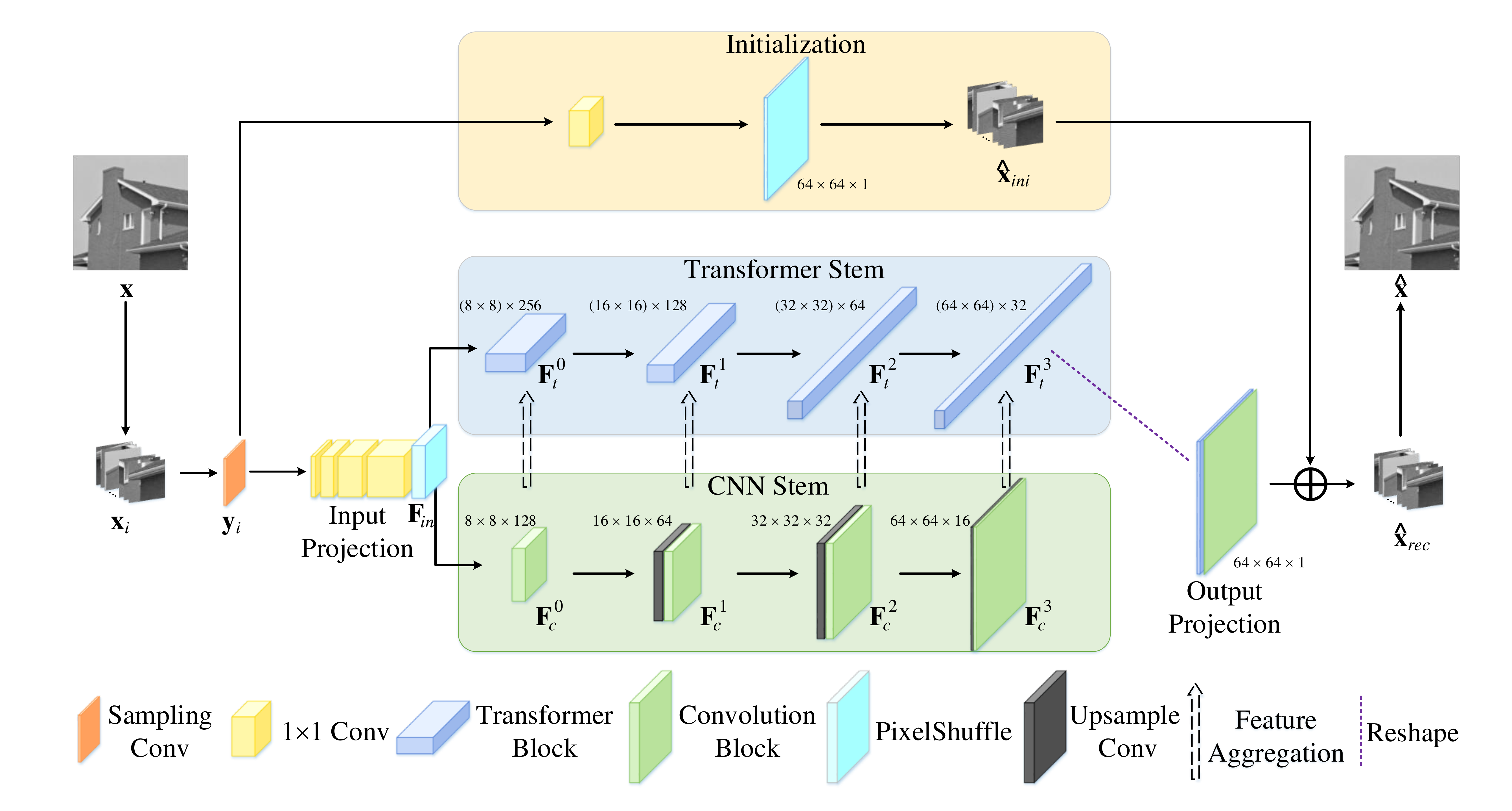}
\caption{Illustration of the pipeline of CSformer. CSformer mainly consists of the initialization, transformer stem and CNN stem. The transformer stem and CNN stem are linked by the feature aggregation.}
\label{fig_pipline}
\end{figure*}

\section{Related work}

In this section, we present the related works. We first review the existing CS methods of natural images in section \ref{rcs}.  Then we provide a brief overview of the recent development of vision transformer in section \ref{rtrans}.

\subsection{Compressive Sensing}
\label{rcs}
CS methods can be classified into two categories: iterative optimization based conventional methods and data-driven based DL methods. Furthermore, we can divide the deep network based approaches into deep unfolding methods and deep straightforward methods. 

\subsubsection{Iterative Optimization based Conventional Methods} The conventional methods mainly rely on sparsity priors to recover the signal from the under-sampled measurements. Some approaches obtain the reconstruction by linear programming based on $L_1$ minimization. Examples of such algorithms involve basis pursuit (BP) \cite{chen2001bp}, least absolute shrinkage and selection operator (LASSO) \cite{tibshirani1996lasso}, the iterative shrinkage/thresholding algorithm (ISTA) \cite{beck2009ista}, and the alternating direction method of multipliers (ADMM) \cite{afonso2010admm}.  In addition, some works improve the recovery performance by exploring image priors. TVAL3 \cite{li2013tval3} utilizes the total variation (TV) regularized to reconstruct images by enhancing the local smoothness. In \cite{metzler2016damp}, D-AMP considers the denoising perspective of the approximate message passing (AMP) \cite{donoho2009message} for CS iterative reconstruction. In general, all the above methods suffer from high computational complexity due to the iterative calculations.

\subsubsection{Deep Unfolding Methods} Deep neural networks have been developed for image CS in the last few years. Deep unfolding methods incorporate the traditional iterative reconstruction and the deep neural networks. Such methods map each iteration into a network layer that preserves the interpretability and performance. Inspired by the D-AMP, Metzler \textit{et al.} \cite{metzler2017learned} implement a learned D-AMP (LDAMP), which unfolds the iterative D-AMP algorithm and combines it with a denoising CNN. In analogy to LDAMP, AMP-Net \cite{zhang2020amp} also applies denoising prior, whereas it has an additional deblocking module and uses a learned sampling matrix.

Moreover, ISTA-Net+ \cite{zhang2018ista} and ISTA-Net++ \cite{you2021ista} design the deep network to mimic the ISTA algorithm for CS reconstruction. The difference is that ISTA-Net++ uses a cross-block learnable sampling strategy and achieves multi-ratio sampling and reconstruction in one model. OPINE-Net \cite{zhang2020optimization} can also be regarded as a variant of ISTA-Net+, except that OPINE-Net simultaneously explores adaptive sampling and recovery. Besides exploring upon on AMP and ISTA, Yang \textit{et al.} \cite{yang2018admm} propose the ADMM-CSNet to reconstruct images with high accuracy and speed by learning the sparse representations, model parameters, and ADMM algorithm from different types of images. The main drawback of the unfolding approaches is that the limitation of parallel training and hardware acceleration owing to its sophisticated and iterative structure.

\subsubsection{Deep Straightforward Methods} Instead of specific priors, the deep straightforward methods directly impose the modeling power of DL free from any constraints. ReconNet \cite{kulkarni2016}, considered as the first deep network based method that brings CNN for CS reconstruction, aims to recover the image from CS measurements via CNN. The reconstruction quality and computational complexity are both superior to the traditional iterative algorithms. Joint learning the sampling with the reconstruction in the whole network further improves the reconstruction performance. Instead of fixing sampling matrix, Shi \textit{et al.} \cite{shi2017deep} implement a convolution layer to replace it and propose a deep network to recover the image named CSNet. In \cite{shi2019image}, they further extend their model to learn binary sampling matrix and bipolar sampling matrix. DR2-Net \cite{yao2019dr2} adopts a fully connected layer to perform the sampling, then stacks several residual learning blocks to improve reconstruction quality. In \cite{sun2020learning}, Sun \textit{et al.} design a 3-D encoder and decoder with the channel attention motivated skip links and introduce the non-local regularization for exploring the long-range dependencies. Sun \textit{et al.} \cite{sun2020dual} propose a dual-path attention network dubbed DPA-Net for CS reconstruction. Two path networks are embedded in the DPA-Net for learning structure and texture, respectively, and then combined by the attention module.

\subsection{Transformer}
\label{rtrans}
The original transformer \cite{vaswani2017attention} is designed for natural language processing (NLP), in which the multi-head self-attention and feed-forward MLP layer excel at handling long-range dependencies of sequence data. Inspired by the power of transformer in NLP, the pioneering work of VIT \cite{dosovitskiy2021an} splits an image into $16\times16$ flattered patches, successfully extending the transformer to image classification task. Swin transformer \cite{Liu_2021_ICCV} designs a hierarchical transformer architecture with the shifted window-based multi-head attentions to reduce the computation cost. Since then, transformer has vaulted into a model on a par with CNN, and the transformer based application of computer vision has mushroomed. Yang \textit{et al.} \cite{yang2020learning} propose a texture transformer network for image super-resolution. They embed the low-resolution image and paired reference image into transformer to obtain a high resolution image. Chen \textit{et al.} \cite{chen2021pre} develop a pre-trained model named image processing transformer (IPT) for several low-level computer vision tasks. They excavate the capability of transformer by using large scale pre-training, and IPT outperforms state-of-the-art methods on super-resolution, denoising, and deraining tasks. Uformer \cite{wang2021uformer} borrows from the structure of U-Net to build transformer to further improve the performance for low-level vision tasks. Liang \textit{et al.} \cite{liang2021swinir} use a
stack of residual swin transformer blocks to achieve state-of-the-art performance on image restoration tasks. In addition, TransGAN proposes a generative adversarial network (GAN)~\cite{goodfellow2014generative, ni2020unpaired, ni2020towards} architecture using pure transformer for image generation. On the other hand, many works aim to combine the strengths from the CNN and transformer effectively. Xie \textit{et al.} \cite{xie2021cort} utilize CNN to extract feature representation and a transformer to model the long-range dependency for 3D medical image segmentation. CoAtNet \cite{dai2021coat} unifies the depthwise convolution and self-attention via a relative attention. Peng \textit{et al.} \cite{Peng_2021_ICCV} propose a dual backbone to combine CNN with visual transformer for visual recognition. ConVit \cite{2021convit} introduces a gated positional self-attention mechanism to bring the convolutional inductive bias to transformer. 


\section{Method}
Fig.~\ref{fig_pipline} illustrates the network architecture of the proposed CSformer for adaptive sampling and reconstruction. The sampling module is applied to sample block by block in the image patches, which are split from the image $\mathbf{x}$ via a non-overlapping way. The sampling matrix is replaced by the learned convolution kernels in each patch. The reconstruction module comprises a linear initialization module, an input projection module, an output projection module, a CNN stem, and a transformer Stem, learning an end-to-end mapping from CS measurements to the recovered images. One stream of the CS measurement is the linear initialization module, including two consecutive operations that a $1\times 1$ convolution and a pixelshuffle layer, to obtain the initial reconstruction $\hat{\mathbf{x}}_{ini}$. The other stream of the CS measurement is to pass through an input projection that contains several layers of $1\times 1$ convolution followed by a pixelshuffle layer to obtain the input feature $\mathbf{F}_{in}$, which matches the input feature sizes for CNN $H_0 \times W_0 \times C_0$ and transformer $(H_0\times W_0)\times C_0$. The trunk recovery network consists of a CNN stem and a transformer Stem.  Each stem contains four blocks with upsample layers to progressively reconstruct features until aligning the patch size. In both branches, convolution features are used to provide local information that complements the features of transformer. The recovery of the trunk recovery network is projected from the transformer output to the single-channel by output projection. CSformer reconstructs the final patches $\hat{\mathbf{x}}_{rec}$ by summing the initial reconstruction and the trunk recovery. Finally, we merge all patches to obtain the final image $\hat{\mathbf{x}}$. 

\begin{figure}[tp]
\centering
\includegraphics[width=0.98\linewidth]{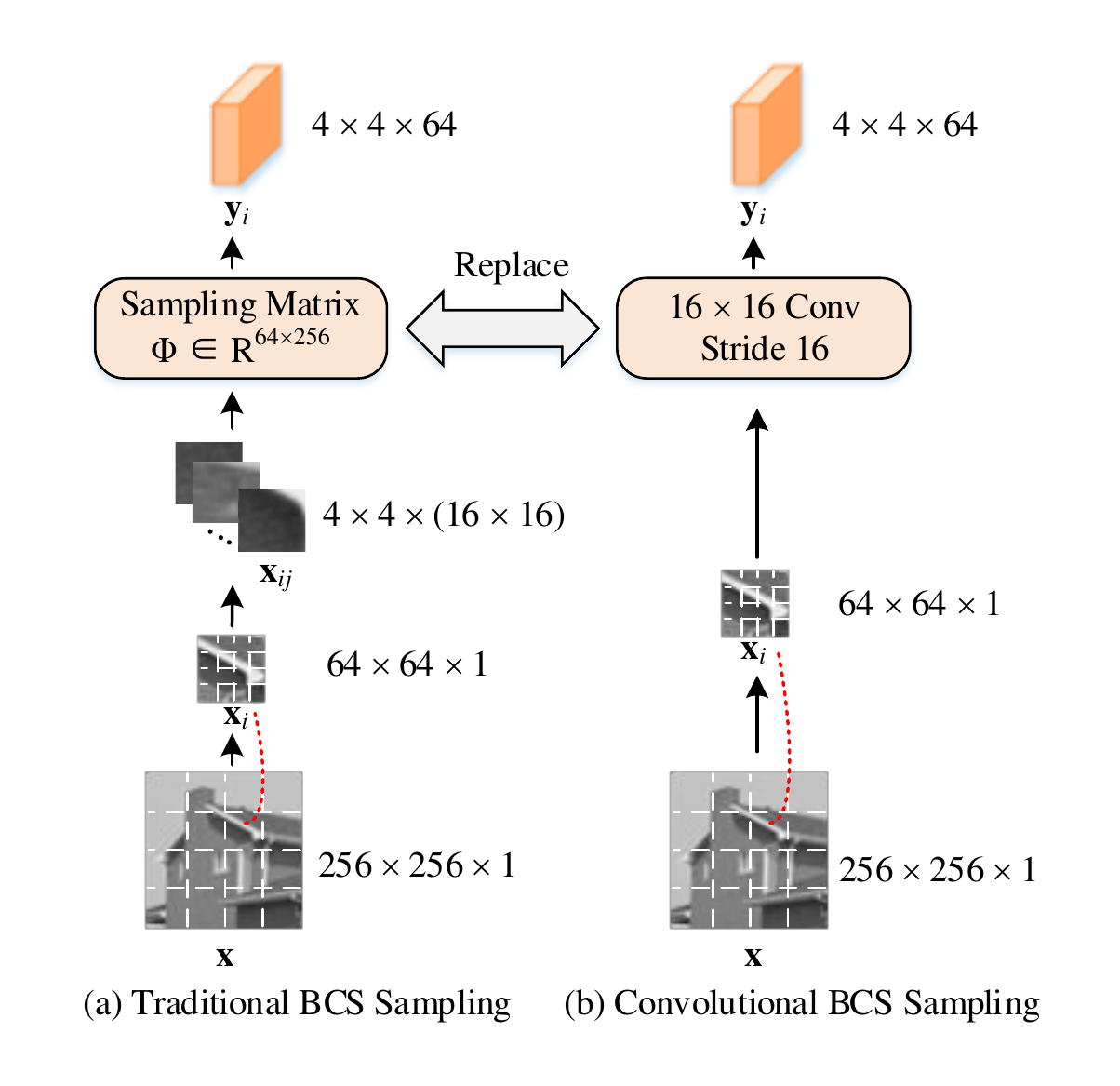}
\caption{Illustration of the details of sampling at CS ratios of 25\%. The traditional block-based CS (BCS) sampling can be equivalently replaced by the convolution.}
\label{fig_sampling}
\end{figure}

\subsection{Sampling}
CSformer samples and reconstructs the whole image by merging the fixed patches. Suppose that $\mathbf{x}_i \in {\rm \mathbb{R}}^{H_p\times W_p\times  1}$ is the patch $i$ of input whole image $\mathbf{x} \in {\rm \mathbb{R}}^{H \times\ W \times 1}$. The sampling operation takes place in patch $\mathbf{x}_i$. We process the block-based CS (BCS) in patch $\mathbf{x}_i$, which decomposes a patch into $B\times B$ non-overlapping blocks. Then the number of blocks is $\frac{H_p}{B}\times\ \frac{W_p}{B}$. Each block is vectorized and subsequently sampled by the measurement matrix $\mathbf{\Phi}$. Suppose that $\mathbf{x}_{ij}$ is the block $j$ of input patch $\mathbf{x}_i$. The corresponding measurement $\mathbf{y}_{ij}$ is obtained by $\mathbf{y}_{ij}=\mathbf{\Phi} \mathbf{x}_{ij}$, where $\mathbf{\Phi} \in {\rm \mathbb{R}}^{m\times B^2}$ and $\frac{m}{B^2}$ represents the sampling ratio. Then the measurement $\mathbf{y}_i\in {\rm \mathbb{R}}^{\frac{H_p}{B} \times \frac{W_p}{B}\times m}$ of the input patch $\mathbf{x}_i$ is obtained by stacking each block. In this paper, the sampling process is replaced by the convolution operation with appropriately sized filters and stride, as shown in Fig.~\ref{fig_sampling}. The sampling convolution can be formulated as:
\begin{equation}
\label{eqn_1}
\mathbf{y}_{ij}=\mathbf{W}_B\otimes \mathbf{x}_{ij},
\end{equation}
where $\mathbf{W}_B$ corresponds to a convolution layer without bias consisting of $m$ filters with $B\times B$ size, and the stride equals to $B$. After applying the convolution operation on the patch $\mathbf{x}_i$, we can obtain the final total CS measurement $\mathbf{y}_i$. As shown in Fig.~\ref{fig_sampling}, the CS measurement $\mathbf{y}_i$ of size $4\times 4 \times 64$ can be acquired from an input patch $\mathbf{x}_i$ of size $64\times 64$ with sampling ratio 0.25 by exploiting a convolution layer using $16$ filters of kernel size  $16\times 16$, stride $= 16$. In this case, $H_p = W_p = 64$, $B = 16$ and $m = 64$. In fact, the adoption of the learned convolutional kernel instead of sampling matrix can efficiently utilize the characteristic of the image, and make the output measurement more easily be used in the following reconstruction module.

\subsection{Initialization}
Given the CS measurements, traditional BCS usually obtain the initial reconstructed block by $\hat{\mathbf{x}}_{ij} = \mathbf{\Phi}^\dagger \mathbf{y}_{ij}$, where $\hat{\mathbf{x}}_{ij}$ is the reconstruction of $\mathbf{x}_{ij}$, and $\mathbf{\Phi}^\dagger \in {\rm \mathbb{R}}^{{B^2}\times m}$ is the pseudo-inverse matrix of $\mathbf{\Phi}$. In CSformer initialization process, we utilze the $1\times 1 \times m$ convolution to replace $\mathbf{\Phi}^\dagger$. The difference is that we can directly implement  the convolution layer on the $\mathbf{y}_i$ to recover the initial patch. The initialization first adopts $B^2$ filters of kernel size $1\times 1 \times m$ to covert the measurement $\mathbf{y}_i$ dimension to $B^2$. Subsequently, the followed pixelshuffle layer is employed to obtain the original patch $\hat{\mathbf{x}}_{i}$. For instance, a measurement with size $4\times 4 \times 64$ is transformed to the initial reconstruction with size $64\times 64 \times 1$ at the CS ratios of $25\%$. In summary, we use the convolution and pixelshuffle to obtain each initial reconstruction, which is a more efficient way as the output is directly a tensor instead of a vector.

\subsection{CNN Stem}
The measurement $\mathbf{y}_i$ is taken as the input of the input projection module that contains several $1\times 1$ convolution layers followed by a pixelshuffle layer to to obtain feature $\mathbf{F}_{in}$ with size $H_0 \times W_0\times C_0$ (by default we set $H_0 = W_0 = 8$). The CNN stem is composed of multiple stages. The first 
stage takes the projected output feature $\mathbf{F}_{in}$ as input. Then the feature passes through the first convolution block to obtain feature $\mathbf{F}_{c}^0$ with size $H_0 \times W_0\times C_0$. Each convolution block is composed of two convolution layers,  followed by a leaky rectified linear unit (ReLU) and a batch norm layer. The kernel size of each convolutional layer is $3 \times 3$ with 1 as the padding size, and the output channel is the same as the input channel. Thus, the resolution and channel size is maintained to be consistent after a convolution block.

To scale up to a higher-resolution feature, we add an upsample module before the rest of convolution block. The upsample convolution module first adopts bicubic upsample to upscale the resolution of the previous feature, and then a $1 \times 1$ convolutional layer is used to reduce the dimension to a half. Thus, the output features of CNN stem  can be represented by $\mathbf{F}_{c}^i \in {\rm \mathbb{R}}^{H_i \times W_i \times C_i}$, where $H_i = 2^{i} \times H_0, W_i = 2^{i} \times W_0, C_i = \frac{C_0}{2^{i}}, i\geq 0$.


\begin{figure}[tp]
\centering
\includegraphics[width=0.98\linewidth]{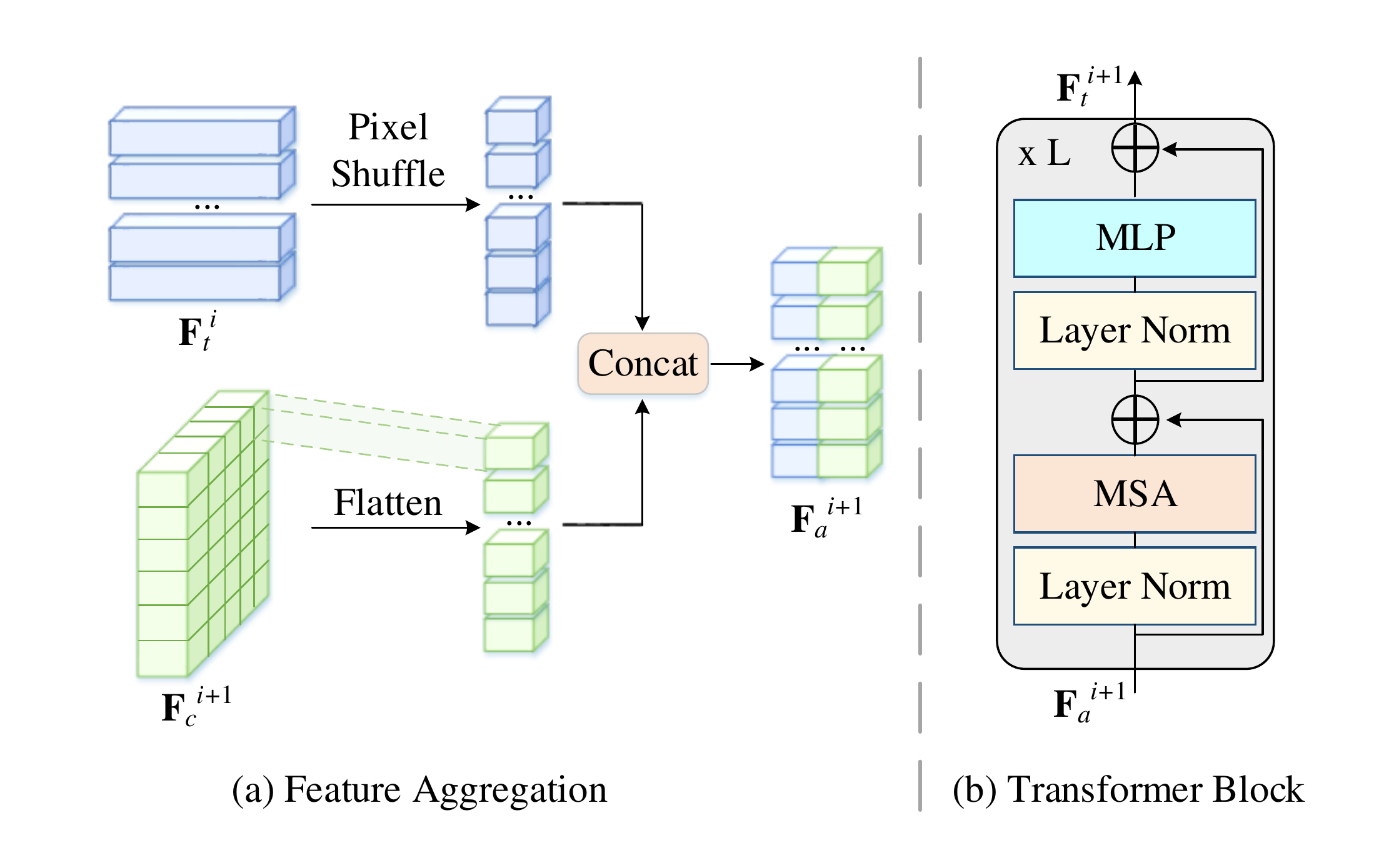}
\caption{Illustration of the implementation details of the transformer stem. (a) Feature aggregation by concatenating the transformer feature $\mathbf{F}_{t}^i$ and CNN feature $\mathbf{F}_{c}^{i+1}$. (b) The detailed transformer block.}
\label{fig_dualstem}
\end{figure}

\subsection{Transformer Stem}
Transformer stem aims to provide further guidance for global restoration with progressive features according to the convolution features. As shown in Fig.~\ref{fig_dualstem}(b), each transformer block stacks $L$ transformer network. The input of transformer is the aggregation feature that bridges the convolution features and transformer features.

\subsubsection{Feature Aggregation}
The aggregation feature fuses the local features from CNN and the global features from transformer via a concatenation way. The feature dimension of CNN stem and transformer stem is inconsistent, such that we need to reshape the CNN features to align with the transformer features. The 2D feature map of CNN with size $H_i \times W_i \times C_i$ needs to be flattened to a 1D sequence $(H_i \times W_i) \times C_i$ for transformer. As can be seen from Fig.~\ref{fig_dualstem}, the aggregation feature is taken as the input to the transformer blocks by concatenating these two features. It is worth mentioning that the input aggregation feature $\mathbf{F}_{a}^0$ of the first transformer block is concatenated by $\mathbf{F}_{in}$ and $\mathbf{F}_{c}^0$. In this way, the first transformer block makes full use of the information in the measurements and introduces local features of CNN. This also aligns with the observation of many studies \cite{Xiao2021early,dai2021coat,raghu2021dovision} that introducing locality in early layers is beneficial for feature representation in transformer. 

After the first transformer block, we obtain the transformer feature $\mathbf{F}_{t}^{0}$ with size $(H_0 \times W_0) \times 2C_0$. The misalignment between the transformer feature with next stage CNN features is further eliminated. We first reshape the 1D sequence of $\mathbf{F}_{t}^{0}$ to 2D feature map with the size $H_0 \times W_0 \times 2C_0$. Subsequently, a pixelshuffle layer is used to upsample the resolution by $2\times$ ratio and reduce the channel dimension to a quarter of the input. We complete the spatial dimension and channel dimension alignment of transformer features and CNN features. Then the aggregation feature is obtained by concatenating the transformer feature and CNN feature. The aggregation feature can be expressed by  $\mathbf{F}_{a}^j \in {\rm \mathbb{R}}^{(H_j \times W_j) \times C_j}$, where $H_j = 2^{j} \times H_0, W_j = 2^{j} \times W_0, C_j = \frac{2C_0}{2^{j}}, j\geq 0$.

\subsubsection{Window-based Transformer}
The standard transformer \cite{vaswani2017attention} takes a series of sequences (tokens) as input and computes self-attention globally between all tokens. However, if we take each pixel as one token in transformer for CS reconstruction, the sequences grow as the resolution increases, resulting in explosive computational complexity for larger resolution. For instance, even a $32\times 32$ image will lead to $1024$ sequences and have $1024^2$ cost of self-attention.
To address the above issue, CSformer performs window-based transformer.  Given an input fusion feature $\mathbf{F}_{a}^j \in {\rm \mathbb{R}}^{(H_j \times W_j ) \times C_j } $ of transformer,  we partition feature into $P \times P$ non-overlapping windows. Then the feature is split into the size of $\frac{H_jW_j}{P^2} \times P^2 \times C_j$, where $\frac{ H_jW_j}{P^2} $ is the total number of windows. 
The multi-head self-attention is computed in each  $P^2$ window. In each window, the feature $\mathbf{F}_{t}^{win} \in {\rm \mathbb{R}}^{P^2  \times \frac{C_j}{h}}$ is computed by the self-attention, where $h$ is the number of heads in the multi-head self attention.  First, the query, key, and value matrices are computed as:
\begin{equation}
\label{eqn_qkv}
\mathbf{Q} = \mathbf{F}_{t}^{win} \times \mathbf{W}_Q, \mathbf{K} = \mathbf{F}_{t}^{win} \times \mathbf{W}_K, \mathbf{V}= \mathbf{F}_{t}^{win} \times \mathbf{W}_V,
\end{equation} 
where $\mathbf{W}_Q$, $\mathbf{W}_K$ and $\mathbf{W}_V$ are the projection matrices with the size $C_j/h \times d$. Subsequently, the self-attention can be formulated by:
\begin{equation}
\label{eqn_att}
\begin{aligned}
&O(\mathbf{F}_{t}^{win}) = \left(\sigma\left(\frac{\mathbf{Q}\mathbf{K}^T}{\sqrt[2]{d}}+\mathbf{E}\right)\right)\mathbf{V},
\end{aligned}
\end{equation} 
where $O(\cdot)$ denotes the self-attention operation, $\sigma(\cdot)$ is the softmax function, and $\mathbf{E}$ is the learnable relative position encoding. The multi-head self-attention is performed for $h$ times self-attention in parallel and concatenates the results to obtain the output. The multi-head self-attention (MSA) based on the windows significantly reduces the computational and GPU memory cost.

Then, the output of MSA passes through a multi-layer perceptron (MLP) consisting of two fully-connected layers with GELU activation for nonlinear transformation. As shown in Fig.~\ref{fig_dualstem}(b), the layer norm $\tau(\cdot)$ is inserted before MSA and MLP and the whole transformer process can be formulated as follows, 
\begin{equation}
\begin{aligned}
&\mathbf{F}_{t}^j = {\rm MSA}(\tau(\mathbf{F}_{a}^j))+ \mathbf{F}_{a}^j, \\
&\mathbf{F}_{t}^j = {\rm MLP}(\tau(\mathbf{F}_{t}^j)) + \mathbf{F}_{t}^j.
\end{aligned}
\end{equation}

After the transformer feature reaches the input resolution $(H_p,W_p)$, the output projection module is used to project the transformer feature to the image space. Before passing through the output projection, we first reshape the transformer feature to a 2D feature. Output projection consists of two convolution layers followed by a tanh action function, which maps the transformer feature to single channel reconstruction patches. Then we sum up the reconstruction patches with the initial reconstruction patches to obtain the final patches $\hat{\mathbf{x}}_{rec}$ and merge all patches to obtain the final reconstructed image $\hat{\mathbf{x}}$.

\subsection{Loss Function}
We optimize the parameters of CSformer by minimizing the the mean
square error (MSE) between the output reconstructed image $\hat{\mathbf{x}}$ and the ground-truth image $\mathbf{x}$ as follows,

\begin{equation}
\mathcal{L}=\left\|\hat{\mathbf{x}}-\mathbf{x}\right\|_2^2.
\end{equation}
It is worth mentioning that the proposed scheme is based on patch reconstruction while the loss function is computed on the whole image. As such, we attenuate the blocking artifacts without other post-processing deblocking modules.

\section{Experiment}
In this section, we first introduce the training settings and evaluation datasets in section \ref{4a}. Section \ref{4c} shows the experimental results of our method compared with state-of-the-art on different test datasets.  Section \ref{4b} analyzes the effectiveness of the proposed approach by comparing the results with those of some variants of CSformer. Section \ref{4d} compares the retraining performance and the computational time.

\begin{figure}[tp]
\centering
\includegraphics[width=0.98\linewidth]{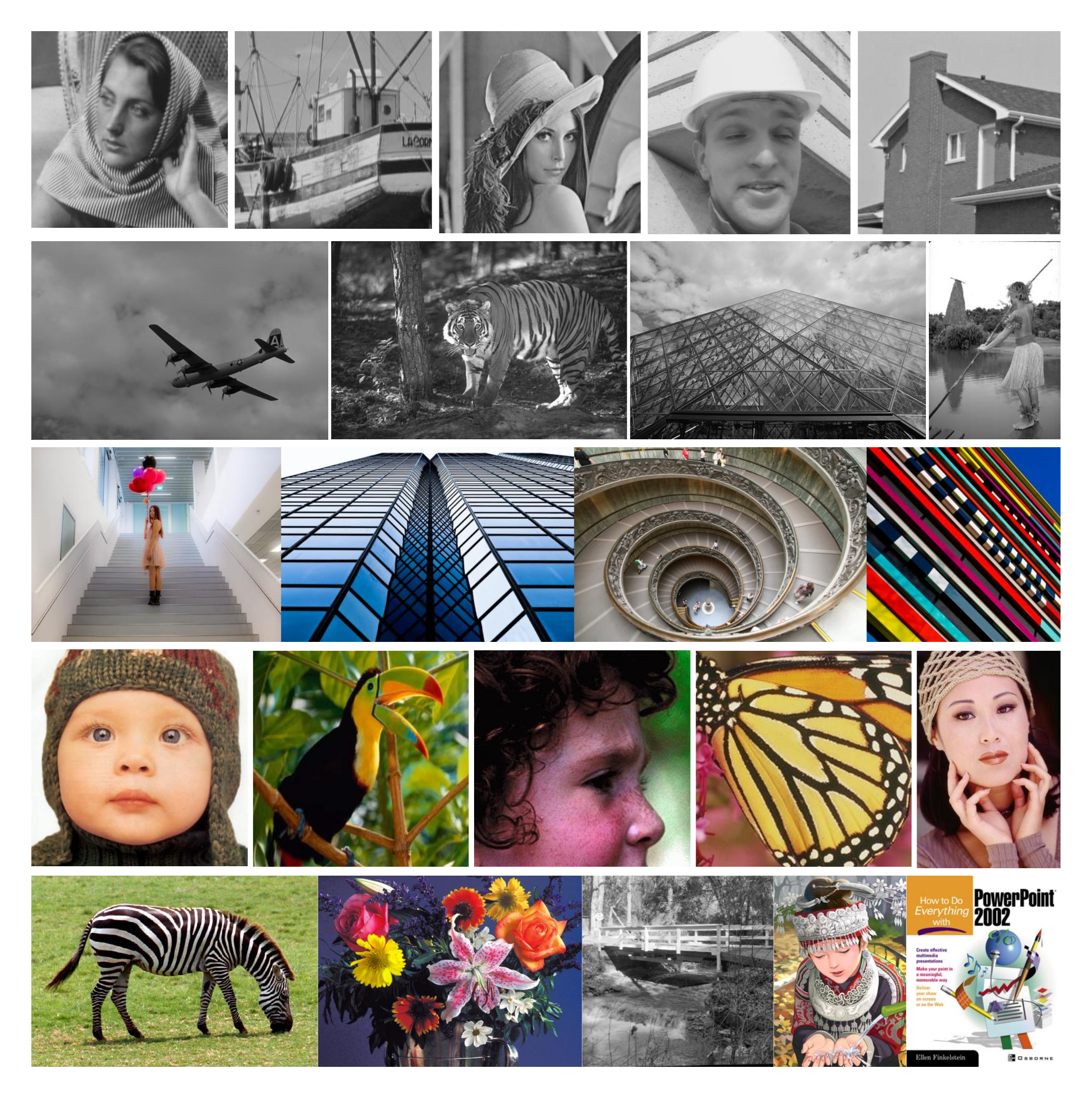}
\caption{Visual samples of various test datasets. From top to bottom, images are from dataset Set11, BSD68, Urban100, Set5, and Set14, respectively.}
\label{fig_dataset}
\end{figure}

\begin{table*}[htp]
\renewcommand\arraystretch{1.2}
\tabcolsep 0.4cm

\centering
\caption{PSNR/SSIM performance Comparisons with different CS ratios on various test datasets. The best one is shown in red and the second best in blue}
\label{tab_mainresults}
\begin{tabular}{c|c|l|l|l|l|l}


\hline
\hline
\multicolumn{1}{l|}{}& \multicolumn{1}{l|}{}&&&&\multicolumn{1}{c|}{}&\\
\multirow{-2}{*}{Dataset}  &
\multirow{-2}{*}{CS ratio} &
\multicolumn{1}{c|}{\multirow{-2}{*}{CSNet$^{+}$}} &
\multicolumn{1}{c|}{\multirow{-2}{*}{DPA-Net}} & \multicolumn{1}{c|}{\multirow{-2}{*}{OPINE-Net}} & \multicolumn{1}{c|}{\multirow{-2}{*}{AMP-Net}} & 
\multicolumn{1}{|c}{\multirow{-2}{*}{CSformer}} \\
\hline

\multirow{5}{*}{Set11} 
& 1\%   &\textcolor{blue}{21.03}/0.5566   & 18.05/0.5011  & 20.15/0.5340  & 20.57/\textcolor{blue}{0.5639}& \textcolor{red}{21.95}/\textcolor{red}{0.6241}\\
& 4\%   &-              & 23.50/0.7205  & \textcolor{blue}{25.69}/\textcolor{blue}{0.7920}  & 25.26/0.7722& \textcolor{red}{26.93}/\textcolor{red}{0.8251}\\
& 10\%  &28.37/0.8580   & 26.99/0.8354  & \textcolor{blue}{29.81}/\textcolor{blue}{0.8884}  & 29.45/0.8787& \textcolor{red}{30.66}/\textcolor{red}{0.9027}\\
& 25\%  &-              & 31.74/0.9238  & \textcolor{blue}{34.86}/\textcolor{blue}{0.9509}  & 34.63/0.9481& \textcolor{red}{35.46}/\textcolor{red}{0.9570}\\
& 50\%  &38.52/0.9749   & 36.73/0.9670  & 40.17/0.9797&\textcolor{blue}{40.34}/\textcolor{blue}{0.9807}
&\textcolor{red}{41.04}/\textcolor{red}{0.9831}\\
\hline

\multirow{5}{*}{BSD68}
& 1\%   &\textcolor{blue}{22.36}/0.5273   & 18.98/0.4643  & 22.11/0.5140  & 22.28/\textcolor{blue}{0.5387}
& \textcolor{red}{23.07}/\textcolor{red}{0.5591}\\
& 4\%   &-              & 23.27/0.6096  & 25.20/\textcolor{blue}{0.6825}  & \textcolor{blue}{25.26}/0.6760 
& \textcolor{red}{25.91}/\textcolor{red}{0.7045}\\
& 10\%  &27.18/0.7766   & 25.57/0.7267  & 27.82/\textcolor{blue}{0.8045}  
& \textcolor{blue}{27.86}/0.7926
& \textcolor{red}{28.28}/\textcolor{red}{0.8078}\\
& 25\%  &-              & 29.68/0.8763  & 31.51/\textcolor{blue}{0.9061}  & \textcolor{blue}{31.74}/0.9048
& \textcolor{red}{31.91}/\textcolor{red}{0.9102}\\
& 50\%  &35.42/0.9614   & 32.89/0.9373  & 36.35/0.9660  & \textcolor{blue}{36.82}/\textcolor{blue}{0.9680} 
& \textcolor{red}{37.16}/\textcolor{red}{0.9714}\\
\hline

\multirow{5}{*}{Urban100}
& 1\%   &20.75/0.5273   & 16.36/0.4162  & 19.82/0.5006 & \textcolor{blue}{20.90}/\textcolor{blue}{0.5328}
& \textcolor{red}{21.94}/\textcolor{red}{0.5885}\\
& 4\%   &-              & 21.64/0.6498  & 23.36/\textcolor{blue}{0.7114}  & \textcolor{blue}{24.15}/0.7029
& \textcolor{red}{26.13}/\textcolor{red}{0.7803}\\
& 10\%  &26.52/0.8053   & 24.54/0.7851  & 26.93/\textcolor{blue}{0.8397}  & \textcolor{blue}{27.38}/0.8270
& \textcolor{red}{29.61}/\textcolor{red}{0.8762}\\
& 25\%  &-              & 28.81/0.8951  & 31.86/\textcolor{blue}{0.9308}  & \textcolor{blue}{32.19}/0.9258
& \textcolor{red}{34.16}/\textcolor{red}{0.9470}\\
& 50\%  &35.25/0.9621   & 32.09/0.9454  & 37.23/\textcolor{blue}{0.9747}  & \textcolor{blue}{37.51}/0.9734
& \textcolor{red}{39.46}/\textcolor{red}{0.9811}\\
\hline

\multirow{5}{*}{Set5}
& 1\%   &\textcolor{blue}{24.18}/0.6478   & 19.02/0.5133  & 21.89/0.6101  & 23.48/\textcolor{blue}{0.6518}
& \textcolor{red}{25.22}/\textcolor{red}{0.7197}\\
& 4\%   &-              & 26.63/0.7767  & 27.95/0.8209  & \textcolor{blue}{29.01}/\textcolor{blue}{0.8359}
& \textcolor{red}{30.31}/\textcolor{red}{0.8686}\\
& 10\%  &32.59/0.9062   & 30.32/0.8713  & 32.51/0.9058  & \textcolor{blue}{33.42}/\textcolor{blue}{0.9140}
& \textcolor{red}{34.20}/\textcolor{red}{0.9262}\\
& 25\%  &-              & 33.96/0.9360  & 36.78/0.9510   
& \textcolor{blue}{38.03}/\textcolor{blue}{0.9586}   
& \textcolor{red}{38.30}/\textcolor{red}{0.9619}\\
& 50\%  &41.79/0.9803  & 39.57/0.9716  
& 41.62/0.9779  
& \textcolor{blue}{42.72}/\textcolor{blue}{0.9818}
& \textcolor{red}{43.55}/\textcolor{red}{0.9845}\\

\hline
\multirow{5}{*}{Set14} 
& 1\%   &\textcolor{blue}{22.92}/0.5630   & 18.30/0.4616  & 21.36/0.5340  & 22.79/\textcolor{blue}{0.5751}
& \textcolor{red}{23.88}/\textcolor{red}{0.6146}\\
& 4\%   &-              & 23.69/0.6534  & 25.50/0.6974  & \textcolor{blue}{26.67}/\textcolor{blue}{0.7219}
& \textcolor{red}{27.78}/\textcolor{red}{0.7581}\\
& 10\%  &29.13/0.8169   & 26.28/0.7693  & 28.77/0.8129  & \textcolor{blue}{29.92}/\textcolor{blue}{0.8312}
& \textcolor{red}{30.85}/\textcolor{red}{0.8515}\\
& 25\%  &-              & 30.15/0.8813  & 33.12/0.9102   & \textcolor{blue}{34.31}/\textcolor{blue}{0.9213}
& \textcolor{red}{35.04}/\textcolor{red}{0.9316}\\
& 50\%  &37.89/0.9631   & 33.78/0.9440  & 38.09/0.9621   & \textcolor{blue}{39.28}/\textcolor{blue}{0.9684}
& \textcolor{red}{40.41}/\textcolor{red}{0.9730}\\

\hline
\hline
\multirow{5}{*}{Direct Average} 
& 1\%
& \textcolor{blue}{22.25}/0.5630 
& 18.14/0.4713
& 21.07/0.5386
& 22.00/\textcolor{blue}{0.5725}
& \textcolor{red}{23.21}/\textcolor{red}{0.6212}\\

& 4\%
& -              
& 23.75/0.6820 
& 25.54/0.7408
& \textcolor{blue}{26.07}/\textcolor{blue}{0.7418}
& \textcolor{red}{27.41}/\textcolor{red}{0.7873}\\

& 10\%
& 28.76/0.8326 
& 26.94/0.7976
& 29.17/\textcolor{blue}{0.8503} 
& \textcolor{blue}{29.61}/0.8487
& \textcolor{red}{30.72}/\textcolor{red}{0.8729}\\

& 25\%
&-              
& 30.87/0.9025 
& 33.63/0.9298
& \textcolor{blue}{34.18}/\textcolor{blue}{0.9317}
& \textcolor{red}{34.97}/\textcolor{red}{0.9415}\\
& 50\%
& 37.77/0.9684   
& 35.01/0.9531 
& 38.69/0.9721 
& \textcolor{blue}{39.33}/\textcolor{blue}{0.9745}
& \textcolor{red}{40.32}/\textcolor{red}{0.9786}\\

\hline
\multirow{5}{*}{Weighted Average} 
& 1\%
& \textcolor{blue}{21.56}/0.5310 
& 17.56/0.4431
& 20.79/0.5122
& 21.55/\textcolor{blue}{0.5426}
& \textcolor{red}{22.55}/\textcolor{red}{0.5855}\\
& 4\%   
&-              
& 22.57/0.6434
& 24.39/\textcolor{blue}{0.7077}
& \textcolor{blue}{24.89}/0.7022
& \textcolor{red}{26.32}/\textcolor{red}{0.7574}\\
& 10\%  
& 27.19/0.8017
& 25.80/0.7689
& 27.67/\textcolor{blue}{0.8301}
& \textcolor{blue}{27.99}/0.8206
& \textcolor{red}{29.42}/\textcolor{red}{0.8537}\\
& 25\%  
&-              
& 29.50/0.8903
& 32.12/\textcolor{blue}{0.9225}
& \textcolor{blue}{32.47}/0.9203
& \textcolor{red}{33.63}/\textcolor{red}{0.9342}\\
& 50\%  
& 35.84/0.9631  
& 32.93/0.9444
& 37.26/0.9712
& \textcolor{blue}{37.69}/\textcolor{blue}{0.9718}
& \textcolor{red}{38.93}/\textcolor{red}{0.9774}\\

\hline
\hline
\end{tabular}
\end{table*}

\subsection{Experimental Settings}
\label{4a}
\subsubsection{Dataset and Metrics}
Training vision transformer is known to be data-hungry. Therefore, we use the COCO 2017 unlabeled images dataset for training, which is a large-scale dataset that consists of over 123K images of high diversity. To reduce the training time, it is worth mentioning that we only use a quarter of the whole training set, \textit{i.e.}, around 40K images for training. We evaluate our method on various widely used benchmark datasets, including Set11 \cite{kulkarni2016}, BSD68 \cite{martin2001database}, Set5 \cite{bevilacqua2012low}, Set14 \cite{zeyde2010single}, Urban100 \cite{huang2015single}. Set11 and BSD68 datasets are composed of 11 and 68 gray images, respectively. Urban100 dataset contains 100 high-resolution challenging city images. Set5 and Set14 datasets have 5 and 14 images with different resolutions. Fig.~\ref{fig_dataset} displays the visual samples of each dataset. We utilize the luminance components of color images for both training and testing. The test images are divided into overlapping patches for testing in the real implementation. The reconstruction results are reported under a range of sampling ratios from 0.1 to 0.5. Peak Signal-to-Noise Ratio (PSNR) and Structural Similarity Index (SSIM) are adopted as the evaluation measures. 

\subsubsection{Training Details}
The training images are cropped into $128\times 128$ images as input, \textit{i.e.}, $H = W = 128$. The size of the fixed patches is $ H_p = W_p = 64$. The sampling convolutional kernel size in the sampling process is set to be $B=16$, \textit{i.e.}, $16\times 16$ convolution layer with stride $=16$. The output feature dimension of input projection  $C_0$ is set to $128$. The window size of window-based multi-head self-attention is set to be $P=8$ for all transformer blocks. Each transformer block stacks $L=5$ transformer network. We use 1 Nvidia 2080Ti card for training our model on Pytorch \cite{paszke2019pytorch}, and the model is optimized by Adam optimizer. The learning rate is initially set as $2\times 10^{-4}$ and the cosine decay strategy is adopted to decrease the learning rate to $1\times 10^{-6}$. The number of iteration is 50,000, and the training time is about 1.5 days.

\begin{figure*}[htp]
\centering
\includegraphics[width=0.98\linewidth]{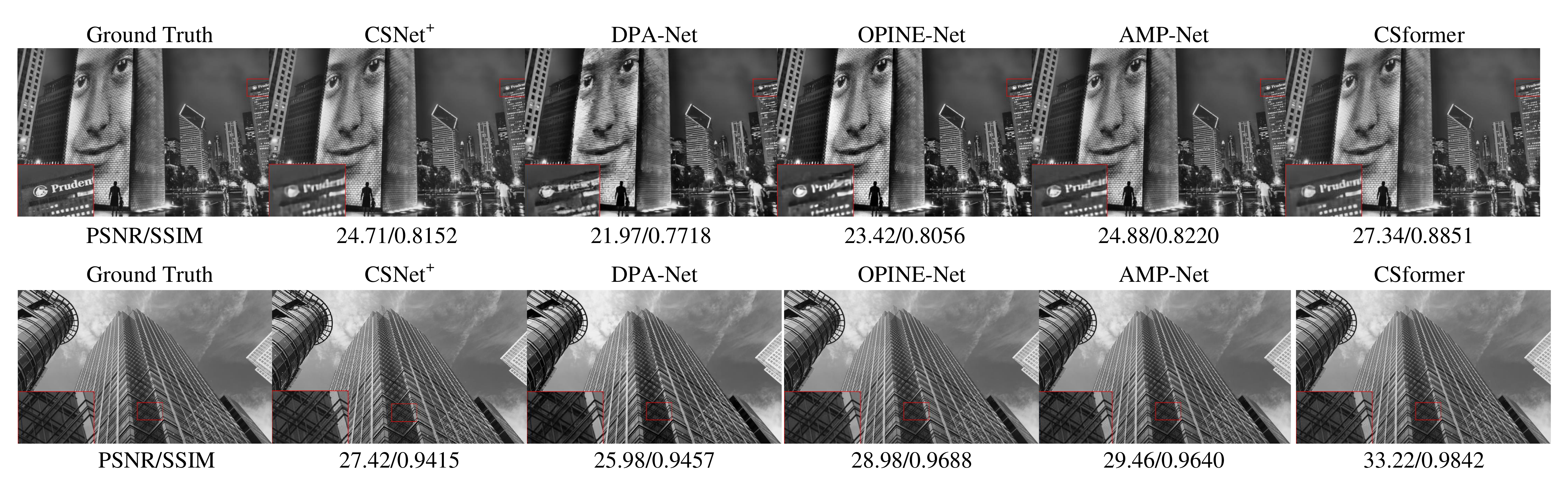}
\caption{Visual quality comparison of various CS methods. The first row is the results at CS ratios of 10\% and the second row is the results at CS ratios of 50\%.}
\label{fig_visualcomparion}
\end{figure*}

\begin{figure*}[htp]
\centering
\includegraphics[width=0.98\linewidth]{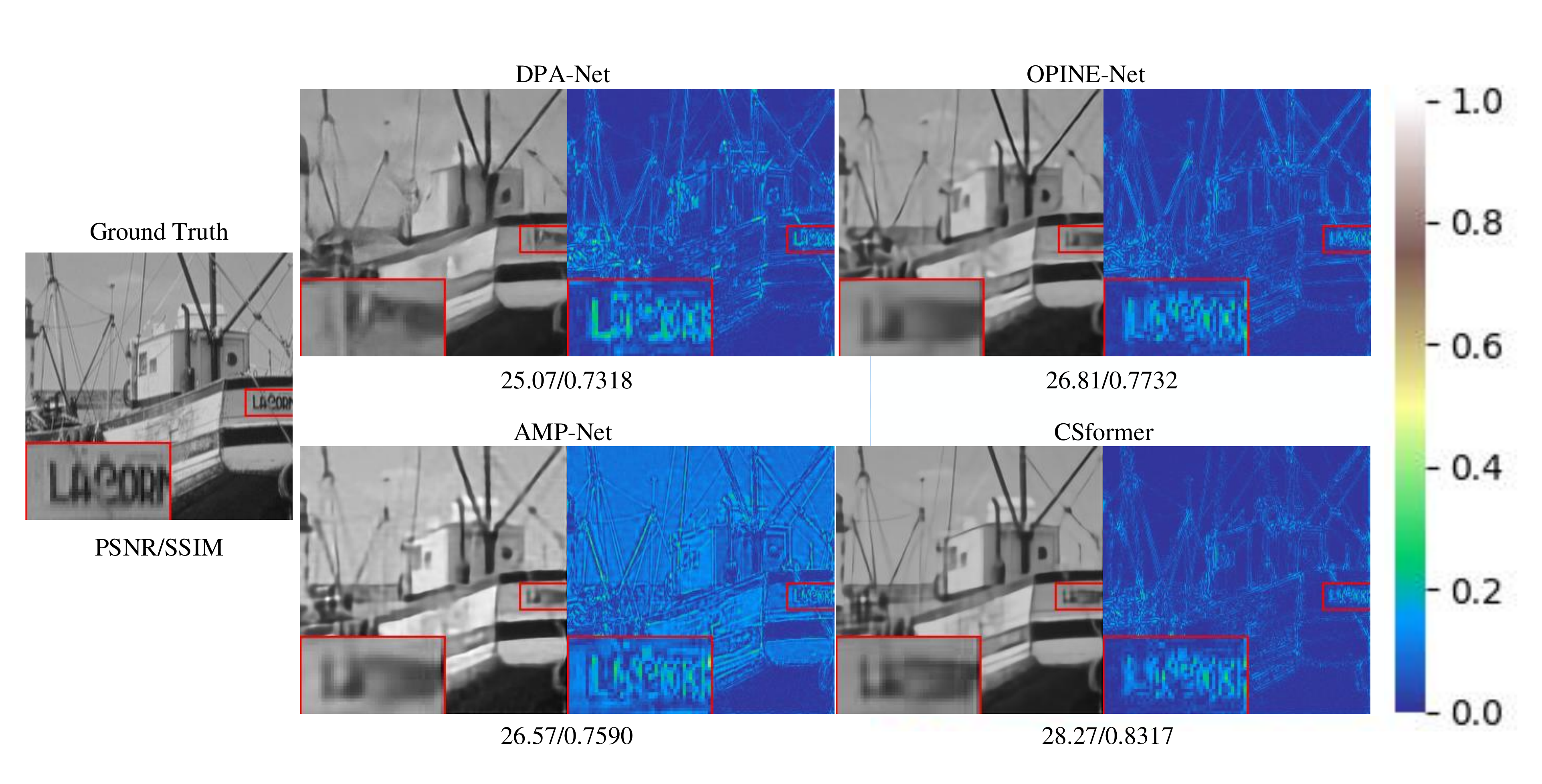}
\caption{Visual quality comparison of the reconstruction image and the absolute residual image at CS ratios of 4\%. The absolute residual intensity map is the result between the recovered image and the ground-truth image.}
\label{fig_visualcomparion_4}
\end{figure*}

\begin{figure*}[htp]
\centering
\includegraphics[width=0.98\linewidth]{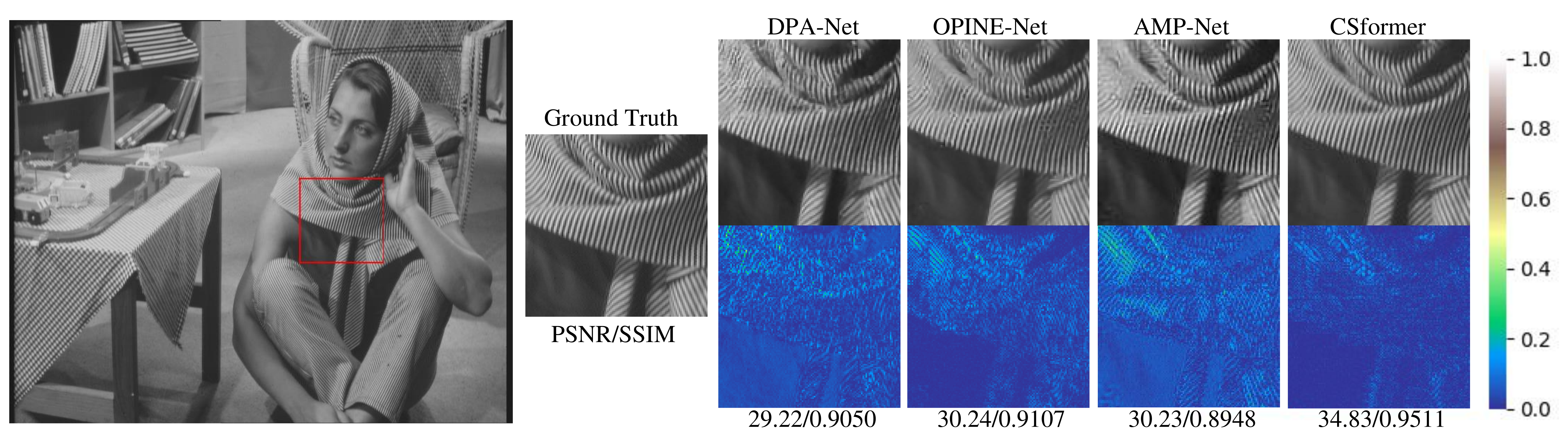}
\caption{Visual quality comparison of the reconstruction image and the absolute residual image at CS ratios of 25\%. The absolute residual intensity map is the result between the recovered image and the ground-truth image.}
\label{fig_visualcomparion_25}
\end{figure*}

\begin{table*}[htp]
\renewcommand\arraystretch{1.2}
\tabcolsep 0.05cm

\centering
\caption{PSNR performance Comparisons with different model sizes. The best results are labeled in bold}
\label{tab_featuredimension}
\begin{tabular}{c|cccc|cccc|cccc|cccc|cccc|c}
\hline
\hline
\multicolumn{1}{c|}{\multirow{2}{*}{Method}} 
& \multicolumn{4}{c|}{Set11}
& \multicolumn{4}{c|}{BSD68} 
& \multicolumn{4}{c|}{Urban100}                                                      
& \multicolumn{4}{c|}{Set5}  
& \multicolumn{4}{c|}{Set14}
&\multicolumn{1}{c}{\multirow{2}{*}{Param}} 
\\ \cline{2-21}
\multicolumn{1}{c|}{}
& \multicolumn{1}{c}{1\%}   
& \multicolumn{1}{c}{10\%}  
& \multicolumn{1}{c}{50\%}  
& Avg.   
& \multicolumn{1}{c}{1\%}   
& \multicolumn{1}{c}{10\%}  
& \multicolumn{1}{c}{50\%}  
& Avg.   
& \multicolumn{1}{c}{1\%}   
& \multicolumn{1}{c}{10\%}  
& \multicolumn{1}{c}{50\%}  
& Avg.    
& \multicolumn{1}{c}{1\%}   
& \multicolumn{1}{c}{10\%}  
& \multicolumn{1}{c}{50\%}  
& Avg.   
& \multicolumn{1}{c}{1\%}   
& \multicolumn{1}{c}{10\%}  
& \multicolumn{1}{c}{50\%}  
& Avg.  
& \multicolumn{1}{c}{}                       
\\ \hline
CSformer$_{64}$                             
& \multicolumn{1}{c}{\textbf{21.99}} 
& \multicolumn{1}{c}{30.26} 
& \multicolumn{1}{c}{40.89}
& 31.05 
& \multicolumn{1}{c}{23.06} 
& \multicolumn{1}{c}{28.14} 
& \multicolumn{1}{c}{37.16} 
& 29.45
& \multicolumn{1}{c}{21.93} 
& \multicolumn{1}{c}{29.06} 
& \multicolumn{1}{c}{38.88} 
& 29.96
& \multicolumn{1}{c}{\textbf{25.24}}
& \multicolumn{1}{c}{33.90} 
& \multicolumn{1}{c}{43.53} 
& 34.22 
& \multicolumn{1}{c}{\textbf{23.90}} 
& \multicolumn{1}{c}{30.56} 
& \multicolumn{1}{c}{40.21} 
& 31.56
& 1.76M                                       
\\
CSformer$_{128}$                               
& \multicolumn{1}{c}{21.95} 
& \multicolumn{1}{c}{30.66} 
& \multicolumn{1}{c}{41.04} 
& 31.22
& \multicolumn{1}{c}{\textbf{23.07}} 
& \multicolumn{1}{c}{28.28} 
& \multicolumn{1}{c}{37.16} 
& 29.50
& \multicolumn{1}{c}{\textbf{21.94}} 
& \multicolumn{1}{c}{29.61} 
& \multicolumn{1}{c}{39.46} 
& 30.34 
& \multicolumn{1}{c}{25.22} 
& \multicolumn{1}{c}{34.20} 
& \multicolumn{1}{c}{43.55}  
& 34.32
& \multicolumn{1}{c}{23.88} 
& \multicolumn{1}{c}{30.85} 
& \multicolumn{1}{c}{40.41} 
& 31.71   
& 6.71M
\\ 
CSformer$_{256}$                              
& \multicolumn{1}{c}{21.94}      
& \multicolumn{1}{c}{\textbf{30.89}}      
& \multicolumn{1}{c}{\textbf{41.22}}      
& \textbf{31.35}  
& \multicolumn{1}{c}{23.03}      
& \multicolumn{1}{c}{\textbf{28.40}}      
& \multicolumn{1}{c}{\textbf{37.26}}      
& \textbf{29.56}      
& \multicolumn{1}{c}{21.85}      
& \multicolumn{1}{c}{\textbf{30.05}}      
& \multicolumn{1}{c}{\textbf{39.75}}      
& \textbf{30.55}       
& \multicolumn{1}{c}{25.18}      
& \multicolumn{1}{c}{\textbf{34.31}}      
& \multicolumn{1}{c}{\textbf{43.76}}      
& \textbf{34.42}      
& \multicolumn{1}{c}{23.84}      
& \multicolumn{1}{c}{\textbf{31.00}}      
& \multicolumn{1}{c}{\textbf{40.56}}      
& \textbf{31.80}     
& 24.94M                                      
\\ 
\hline
\hline
\end{tabular}
\end{table*}

\subsection{Performance Comparisons}
\label{4c}
To facilitate comparisons, we evaluate the performance of our CSformer on five widely used testsets, and compare our method with four recent representatives DL based CS state-of-the-art methods, including CSNet$^+$ \cite{shi2019image}, DPA-Net \cite{sun2020dual}, OPINE-Net \cite{zhang2020optimization} and AMP-Net \cite{zhang2020amp}. The results of other methods are obtained by their public pre-trained model.

To display the comprehensive performance comparisons over multiple datasets, we utilize two commonly-used average measures to evaluate the average performance over the five test databases, as suggested in \cite{ni2018gabor}. The two average measures can be defined as follows:
\begin{equation}
\begin{aligned}
\overline{s} = \frac{\sum\nolimits_{i=1}^{D}s_i\cdot \beta_i}{\sum\nolimits_{i=1}^{D}\beta_i},
\end{aligned}
\end{equation}
where $D$ denotes the total number of databaset ($D = 5$ in this paper), $s_i$ represents the value of the performance index (e.g. PSNR, SSIM) on the $i$-th dataset, and $\beta_i$ is the corresponding weight on the $i$-th dataset. The first average measurement is \textit{Direct Average} with $\beta_i = 1$. The second average measurement is \textit{Weighted Average}, where $\beta_i$ is set as the number of images in the $i$-th dataset (e.g. 11 for the Set11 dataset, 100 for the Urban100 dataset).

Table \ref{tab_mainresults} shows the average PSNR and SSIM performance of different methods at different CS ratios across all five datasets. It can be obviously observed that the proposed CSformer achieves the both highest PSNR and SSIM results for different ratios on all datasets. Our approach achieves a large gap (1$ \sim $2 dB) across all CS ratios in Urban100 dataset that contains more images with larger resolution. The Direct Average and Weighted Average show our proposed CSformer outperforms all state-of-the-art models
under comparison. The improvement of performance is mainly attributed to the powerful feature representation ability by bridging the two strong neural networks, CNN and transformer. Experimental results demonstrate that CSformer has better generalization ability and recovery ability for limit sampling under the premise that all sampling rates can achieve optimal performance.

In Fig.~\ref{fig_visualcomparion}, we show the reconstructed images of all the methods at CS ratios of 10$\%$ and 50$\%$. The proposed CSformer is able to recover more fine detail and more clear edges than other methods. Fig.~\ref{fig_visualcomparion_4} shows the qualitative comparison of the reconstruction image and the absolute residual intensity map with different methods at CS ratios of 4\%. The absolute residual intensity map is the intensity map of the absolute residual between the recovered image and the ground-truth image. As shown in Fig.~\ref{fig_visualcomparion_4}, our CSformer can recover more fine details and structure due to the help of CNN to transformer. Compared to DPA-Net, which uses the dual-Path CNN structure, our clarity is significantly improved. Compared to the deep unfolding methods OPINE-Net and AMP-Net, our CSformer reduces the artifact and provides more reasonable reconstruction. The visual quality results at CS ratio of 25\% are shown in Fig.~\ref{fig_visualcomparion_25}. The improvement can be seen more clearly in the residual map that the reconstructed texture detail of our approach is finer. The visual quality comparisons clearly demonstrate the effectiveness of the proposed CSformer. Overall, the quantitative and qualitative comparisons with several competing methods verify the superiority of CSformer.

\subsection{Ablation Studies}
\label{4b}
This subsection first presents the ablation studies on the feature dimension and feature aggregation. Subsequently, network structure is analyzed to investigate the effects of the dual structure in our CSformer. Moreover, we visualize the feature map and the feature similarity to verify that our hybrid framework effectively bridges CNN and transformer.

\subsubsection{Feature Dimension}
Table \ref{tab_featuredimension} shows the results for different dimensions, where the subscript represents the dimension of $C_0$. The smaller CSformer$_{64}$ is capable of achieving good performance on the five datasets. The CSformer$_{128}$ outperforms CSformer$_{64}$ at most of CS ratios. The largest improvement appears on the Urban100 dataset with average 0.4 dB. In addition, there are about 0.2 dB PSNR gains over Set11 and Set14. The larger CSformer$_{256}$ achieves around 0.1$ \sim $0.2 dB gains than the second one but has the maximum number of parameters. To balance the performance and model size, we adopt $C_0 = 128$ for our CSformer by default.

\begin{figure}[tp]
\centering
\includegraphics[width=0.98\linewidth]{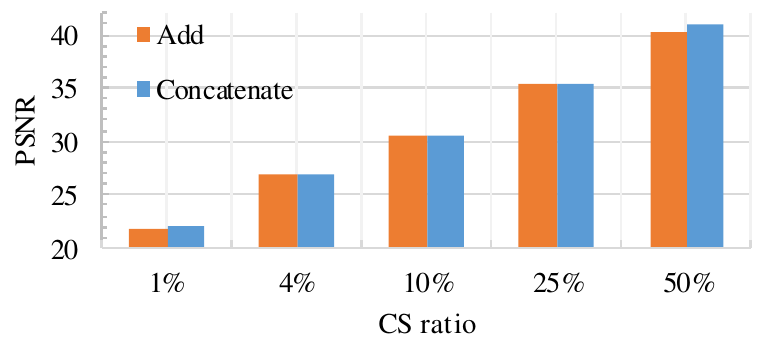}
\caption{Performance comparison between the adding feature fusion and concatenating feature aggregation on Set11.}
\label{fig_add_set11}
\end{figure}

\begin{figure}[tp]
\centering
\includegraphics[width=0.98\linewidth]{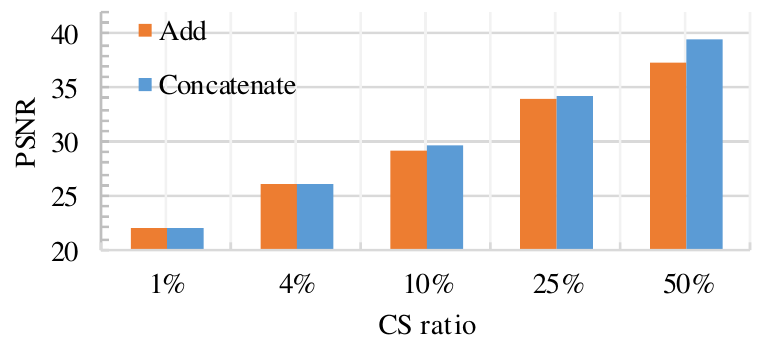}
\caption{Performance comparison between the adding feature fusion and concatenating feature aggregation on Urban100.}
\label{fig_add_urban100}
\end{figure}

\begin{figure*}[htp]
\centering
\includegraphics[width=0.98\linewidth]{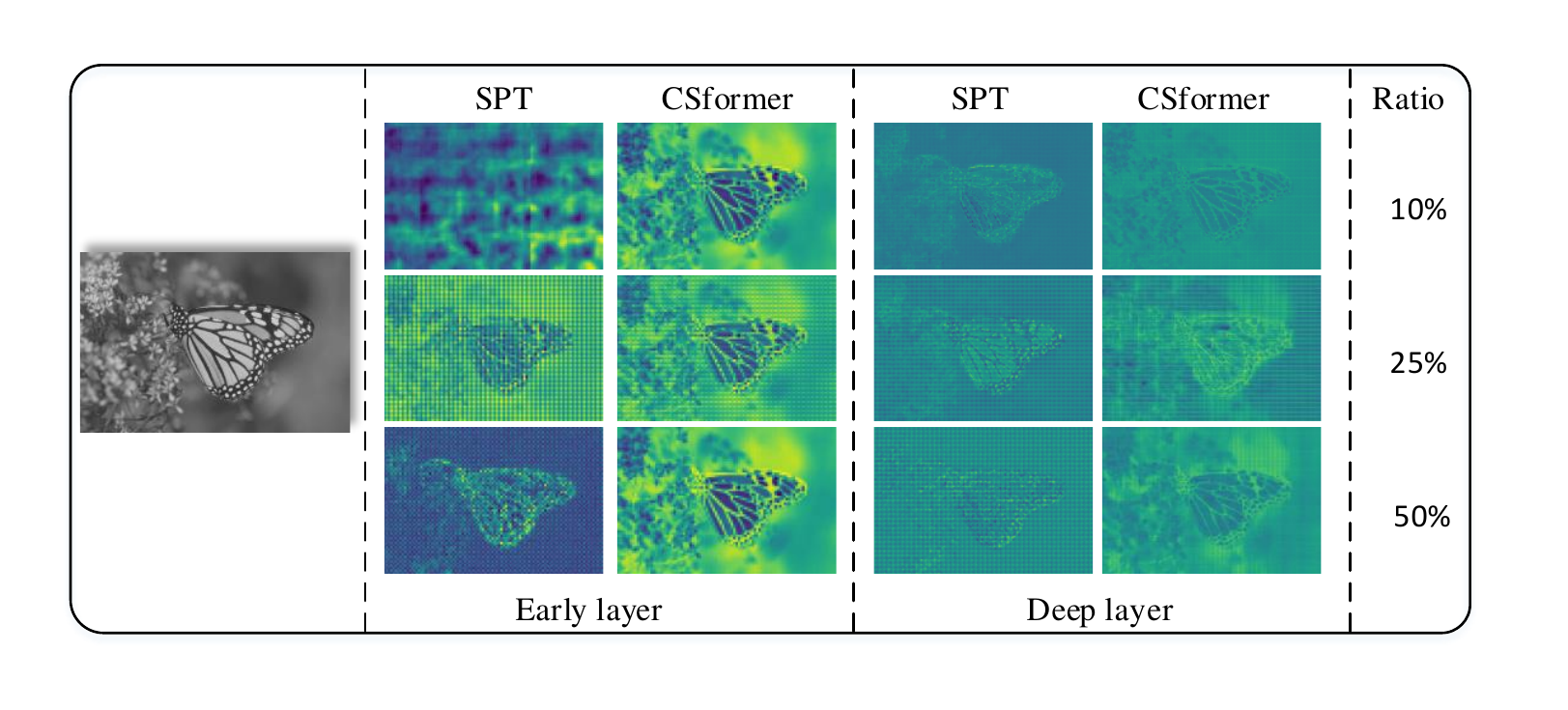}
\caption{Comparison of feature map of the single-path transformer (SPT) and the proposed CSformer.  
The SPT tends to activate more global areas than the local region, while CSformer enhances the locality of features.}
\label{fig_visualfeatures}
\end{figure*}

\subsubsection{Feature Aggregation}
CSformer adopts concatenation operator to aggregate features from different stems. To illustrate the effectiveness of this way, we construct a variant that the CNN features and transformer features are added rather than concatenated. It is worth mentioning that replacing the concatenation operation directly with the addition operation will cause the feature dimension of the transformer block to be halved. Thus, for a fair comparison, we modify the output dimension of CNN block to keep the input dimension of the transformer stem unchanged for using the adding fusion way. The parameters of the CSformer using adding feature fusion are 9.04 M, and the parameters of the CSformer using concatenating feature aggregation are 6.71 M. Fig.~\ref{fig_add_set11} shows the PSNR results on the Set11 dataset. The concatenating feature aggregation shows superior PSNR performance with different samling ratios and has fewer parameters. The adding feature fusion operation achieves a close performance when CS ratios are less than 50\%. The gap is most obvious at CS ratios of 50\%, which shows the concatenation way can make better use of the complementarity of the CNN features and transformer features at higher sampling ratios. The same pattern is observed on the Urban100 dataset, as shown in Fig.~\ref{fig_add_urban100}. The default concatenation aggregation way has superior performance compared to the way of adding fusion on Urban100. The improvement is up to 2.06 dB at 50\% CS ratios and around 0.1$\sim$0.3 dB at 1\% to 25\%.

\begin{table}[tp]
\renewcommand\arraystretch{1.2}
\tabcolsep 0.2cm
\centering
\caption{PSNR performance comparison of the proposed CSformer with the single-path transformer (SPT) on Set11}
\label{table_spt_set11}
\begin{tabular}{c|c|c|c|c|c|c}
\hline
\hline
Method   & 1\%   & 4\%   & 10\%  & 25\%  & 50\%  & Param \\
\hline
SPT      & 21.71 & 26.95 & 30.76 & 35.50 & 41.05 & 7.40M \\
CSformer & 21.95 & 26.93 & 30.66 & 35.46 & 41.04 & 6.71M \\
\hline
\hline
\end{tabular}
\end{table}

\begin{table}[tp]
\renewcommand\arraystretch{1.2}
\tabcolsep 0.2cm
\centering
\caption{PSNR performance comparison of the proposed CSformer with the single-path transformer (SPT) on Urban100}
\label{table_spt_Urban}
\begin{tabular}{c|c|c|c|c|c|c}
\hline
\hline
Method   & 1\%   & 4\%   & 10\%  & 25\%  & 50\%  & Param \\
\hline
SPT      & 21.81 & 25.91 & 29.50 & 33.61 & 38.62 & 7.40M \\
CSformer & 21.94 & 26.13 & 29.61 & 34.16 & 39.46 & 6.71M \\
\hline
\hline
\end{tabular}
\end{table}

\subsubsection{Dual Stem}
CSformer is a dual stems model, aiming to couple the efficiency of  convolution in extracting local features with the power of transformer in modeling global representations. To evaluate the benefits of these two branches, we build a single-path model, named “SPT”, which only uses transformer for reconstruction. For a fair comparison, we add one more $1 \times 1$ convolution before transformer block and set $C_0 = 256$ to maintain the consistency of resolution and dimension in transformer block while keeping all others unchanged. The testing is implemented on the Set11 dataset and Urban100 dataset as depicted in Table~\ref{table_spt_set11} and Table \ref{table_spt_Urban}. CSformer shows a better result on the Set11 dataset at CS ratios of 1\% while has slight performances drop than SPT at other ratios. This is partly due to the increase in the number of parameters and partly reflects the powerful modeling capability of the transformer network. On the Urban100 dataset, CSformer shows superior PSNR performance at different CS ratios with at most 0.84 dB gains. The gap between these two methods ascends with the increase of sampling ratio and achieves the largest gap at CS ratio of 50\%. The improvement of CSformer is more noticeable at high ratios. The reason can be explained by the fact that the trunk recovery network recovers the residuals according to the initial reconstruction, while under high sampling ratios the initial reconstruction is relatively sufficient. Therefore, the detailed and local information provided by CNN is more helpful for the final reconstruction. Meanwhile, CSformer plays more critical roles on the Urban100 dataset than the Set11 dataset. The reason can be 
attributed to the fact that the Urban100 dataset has more textured data, making the local information more helpful for the reconstruction. In this case, the convolution operation is more efficient and practical for image local feature extraction.

\begin{figure}[tp]
\centering
\includegraphics[width=0.98\linewidth]{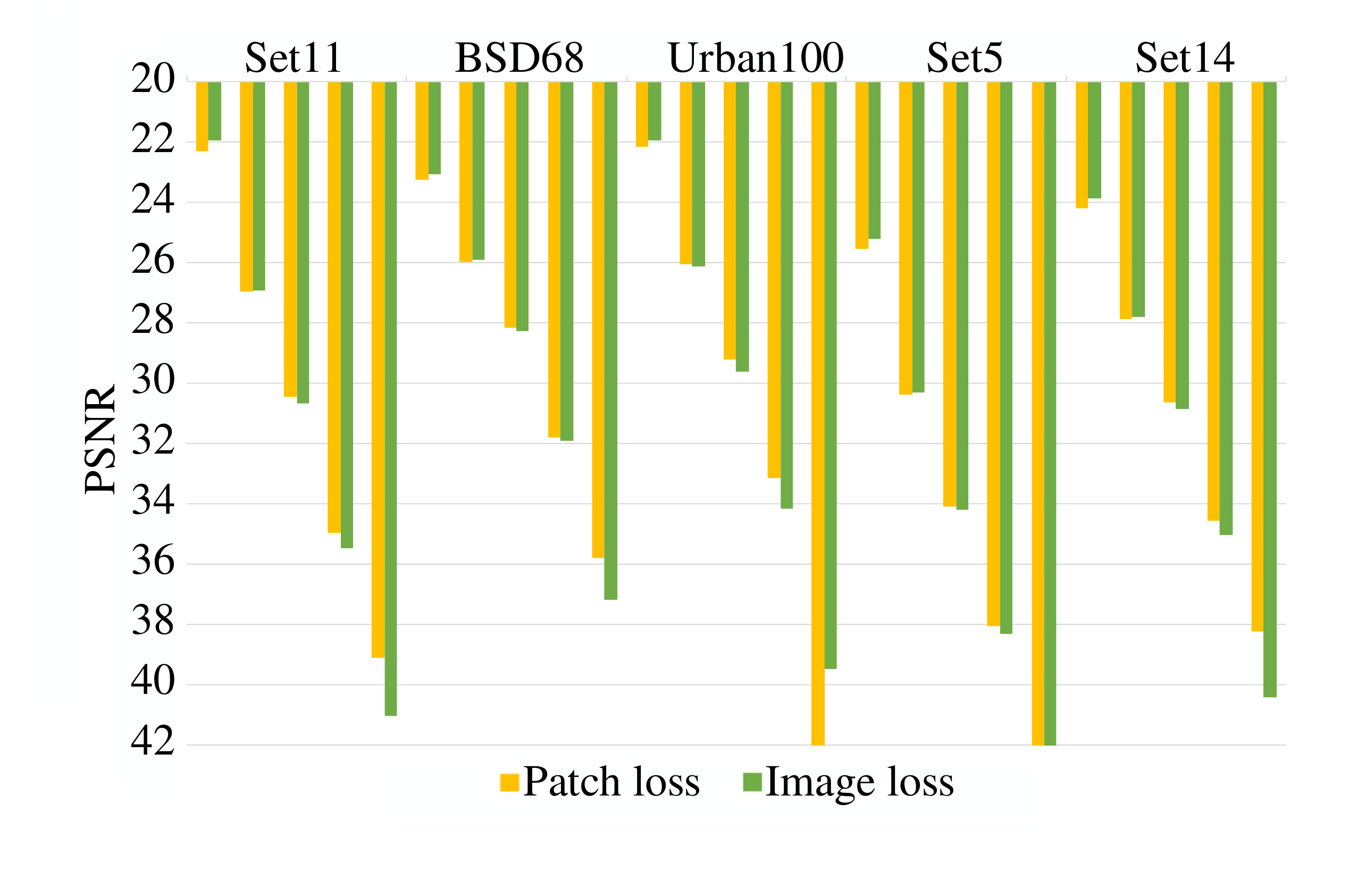}
\caption{Comparison between calculating loss on the patch (patch loss) and on the merged image (image loss). Within each dataset, the sampling ratios from left to right were 1\%, 4\%, 10\%, 25\%, and 50\%, respectively.}
\label{fig_loss}
\end{figure}

\subsubsection{Image Loss}
The proposed CSformer relies on the loss between images instead of patches to reduce the blocking artifact by merging the output patches to the image. In Fig.~\ref{fig_loss}, we compare the image loss with another version of CSformer that calculates the loss between the input patches and output patches. Through experiments on five testing datasets, it is found that image loss adopted by CSformer can significantly improve performance without additional post-processing modules, especially in the case of higher sampling rates.

\subsubsection{Feature Analysis}
We investigate the difference of the internal features representations between CNN and transformer by feature visualization and feature similarity. In the first analysis, we visualize the feature maps in Fig.~\ref{fig_visualfeatures}.  It can be seen that compared with the CSformer, the SPT tends to activate more global areas than the local region. Besides, with the help of the local information extracted by CNN, the detailed textures are remained in the CSformer compared to SPT. This figure shows the ability of CSformer in bridging the local feature and global representation, which enhances the locality of features through convolution starting from the early layer. The early local intervention is a helpful complement to transformer features.

\begin{figure}[tp]
\centering
\includegraphics[width=0.98\linewidth]{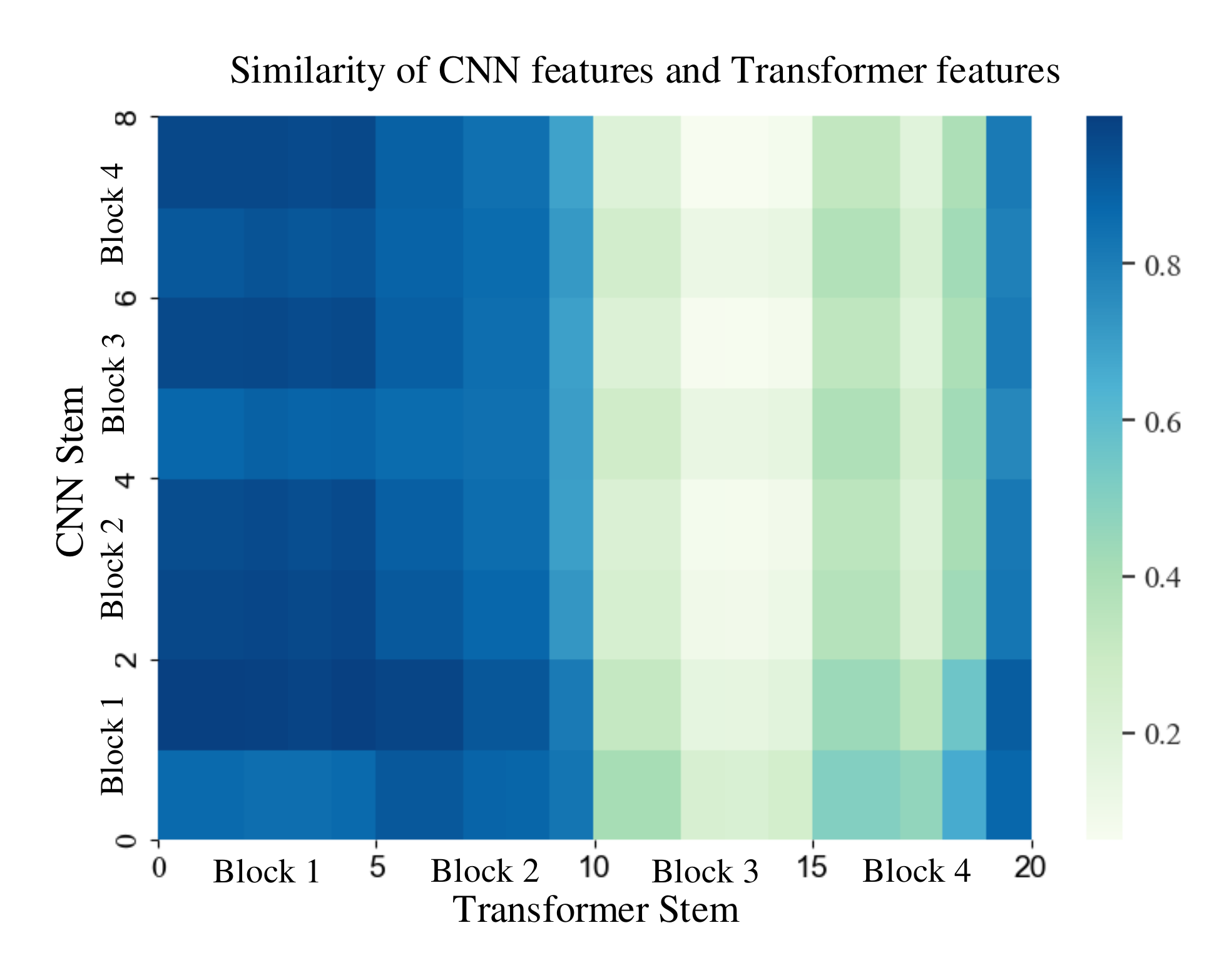}
\caption{Comparison of similarity of CNN stem and transformer Stem. The lower layers of transformer stem are similar to the deep layers of CNN stem. The middle layers show dissimilarity, and the deep layers shows moderate similarity.}
\label{fig_similarityfeatures}
\end{figure}

In Fig.~\ref{fig_similarityfeatures}, we extract the CNN features and transformer features from CNN stem and transformer stem, respectively. We analyze the features from the perspective of representation similarity using centered kernel alignment \cite{kornblith2019similarity}. It is worth mentioning that the transformer features already contain the CNN features as the fusion features the input of transformer block. We observe the lower layers of transformer block are similar to the deep layers of CNN. It shows the transformer has a good ability to capture the long-dependence from the beginning, while CNN requires more dependence on the stacking of layers to enhance long-distance feature dependencies. In addition, it indicates that the CNN features play a more critical role in the early layers than deep layers. The middle features show weak similarity, which indicates the transformer features show more dominant effects. The deep layers show moderate similarity, and it illustrates our CSformer balances the local and global representation in the deep layers.

\begin{table}[tp]
\renewcommand\arraystretch{1.1}
\tabcolsep 0.1cm

\centering
\caption{PSNR performance retraining on the COCO dataset. The best results are labeled in bold and the gap form the best performance is indicated in parentheses}
\label{tab_retrain}
\begin{tabular}{c|c|c|c|c}

\hline
\hline
\multicolumn{1}{l|}{}
& \multicolumn{1}{l|}{}
&
&\multicolumn{1}{c|}{}
&\\
\multirow{-2}{*}{Dataset}  &
\multirow{-2}{*}{CS ratio} &
\multicolumn{1}{c|}{\multirow{-2}{*}{OPINE-Net}} & \multicolumn{1}{c|}{\multirow{-2}{*}{AMP-Net}} & 
\multicolumn{1}{|c}{\multirow{-2}{*}{CSformer}} \\
\hline

\multirow{5}{*}{Set11} 
& 1\%  
& 20.30
& 20.09
& \textbf{21.95}\\
& 4\%  
& 25.58
& 24.93
& \textbf{26.93}\\
& 10\%  
& 29.90 
& 29.51
& \textbf{30.66}\\
& 25\%  
& 34.66
& 34.47
& \textbf{35.46}\\
& 50\%  
& 39.56
& 39.99
&\textbf{41.04}\\
\hline

\multirow{5}{*}{BSD68} 
& 1\%  
& 21.98
& 21.88
& \textbf{23.07}\\
& 4\%  
& 25.22
& 24.97
& \textbf{25.91}\\
& 10\%  
& 27.87
& 27.64
& \textbf{28.28}\\
& 25\%  
& 31.53
& 31.40
& \textbf{31.91}\\
& 50\%  
& 36.16
& 36.30
&\textbf{37.16}\\
\hline

\multirow{5}{*}{Urban100} 
& 1\%  
& 20.91
& 20.91
& \textbf{21.94}\\
& 4\%  
& 24.52
& 24.67
& \textbf{26.13}\\
& 10\%  
& 28.72
& 28.03
& \textbf{29.61}\\
& 25\%  
& 33.26
& 32.93
& \textbf{34.16}\\
& 50\%  
& 38.10
& 38.63
&\textbf{39.46}\\
\hline

\multirow{5}{*}{Set5} 
& 1\%  
& 23.23
& 23.49
& \textbf{25.22}\\
& 4\%  
& 28.96
& 28.58
& \textbf{30.31}\\
& 10\%  
& 33.48
& 33.21
& \textbf{34.20}\\
& 25\%  
& 37.71
& 37.72
& \textbf{38.30}\\
& 50\%  
& 42.12
& 42.54
&\textbf{43.55}\\
\hline

\multirow{5}{*}{Set14} 
& 1\%  
& 22.58
& 22.71
& \textbf{23.88}\\
& 4\%  
& 26.83
& 26.86
& \textbf{27.78}\\
& 10\%  
& 30.26
& 30.16
& \textbf{30.85}\\
& 25\%  
& 34.42
& 34.36
& \textbf{35.04}\\
& 50\%  
& 39.04
& 39.45
&\textbf{40.41}\\
\hline
\hline

\multirow{5}{*}{Direct Average} 
& 1\%  
& 21.80 (\textcolor{red}{-1.41})
& 21.82 (\textcolor{red}{-1.39})
& \textbf{23.21}\\
& 4\%  
& 26.22 (\textcolor{red}{-1.19})
& 26.00 (\textcolor{red}{-1.41})
& \textbf{27.41}\\
& 10\%  
& 30.05 (\textcolor{red}{-0.67}) 
& 29.71 (\textcolor{red}{-1.01})
& \textbf{30.72}\\
& 25\%  
& 34.32 (\textcolor{red}{-0.65}) 
& 34.18 (\textcolor{red}{-0.79}) 
& \textbf{34.97}\\
& 50\%  
& 39.00 (\textcolor{red}{-1.32}) 
& 39.38 (\textcolor{red}{-0.94}) 
& \textbf{40.32}\\
\hline

\multirow{5}{*}{Weighted Average} 
& 1\%  
& 21.42 (\textcolor{red}{-1.13}) 
& 21.39 (\textcolor{red}{-1.16}) 
& \textbf{22.55}\\
& 4\%  
& 25.09 (\textcolor{red}{-1.23}) 
& 25.04 (\textcolor{red}{-1.28}) 
& \textbf{26.32}\\
& 10\%  
& 28.72 (\textcolor{red}{-0.70}) 
& 28.26 (\textcolor{red}{-1.16}) 
& \textbf{29.42}\\
& 25\%  
& 32.94 (\textcolor{red}{-0.69}) 
& 32.71 (\textcolor{red}{-0.92}) 
& \textbf{33.63}\\
& 50\%  
& 37.68 (\textcolor{red}{-1.25}) 
& 38.06 (\textcolor{red}{-0.87}) 
&\textbf{38.93}\\

\hline
\hline
\end{tabular}
\end{table}

\begin{table}[tp]
\renewcommand\arraystretch{1.2}
\tabcolsep 0.05cm

\centering
\caption{Comparison of the model size and running time(in seconds) for reconstructing a $256 \times 256$ image}
\label{table_sizeandtime}
\begin{tabular}{c|c|c|c|c|c}
\hline
\hline
Method & CSNet$^+$  & DPA-Net &OPINE-Net  &AMP-Net & CSformer \\ 
\hline
Param  & 5.00M   & 9.78M   & 1.10M   & 1.53M    & 6.71M     \\ 
Time   & 0.0176 & 0.0339  & 0.0140  & 0.0322   & 5.0765    \\ \hline
\hline
\end{tabular}
\end{table}

\subsection{Analysis on the Retraining Performance and Running Time}
\label{4d}
We retrain the AMP-Net and OPINE-Net on the COCO dataset to show their performance on the larger training dataset in Table~\ref{tab_retrain}. The original AMP-Net is trained on the BSD500 dataset \cite{arbelaez2010contour}, and OPINE-Net is trained on the T91 dataset \cite{kulkarni2016}. As shown in Table~\ref{tab_retrain}, the CSformer achieves the highest PSNR results under the same training dataset. Compared with the model trained on the BSD500 dataset and T91 dataset, the performances of the other two methods show varying degrees of improvement or decline across multiple datasets.

Table~\ref{table_sizeandtime} provides the parameter number of various CS methods at CS ratio of 50\% and the time consuming analysis for reconstructing a $256 \times 256$ image. Considering that we utilize the transformer model and CNN model, the total parameters of our method are still 30\% lower than the DPA-Net using the dual-path CNN structure. Though the running time increases, our proposed CSformer achieves the best performance and generalization capabilities.

\section{Conclusion}
In this paper, we propose a novel dual-stem network named CSformer, which bridges the CNN and transformer networks for adaptive sampling and reconstruction of CS. The sampling stage adaptively learns the sampling matrix and adopts sampling block by block. In the reconstruction stage, we design a dual-stem structure to combine the two types of features and gradually increase feature resolution to reduce memory cost and computation complexity. Experiments show that our CSformer effectively utilizes the complementary of transformer and CNN, outperforming the pure single-path transformer. The proposed CSformer achieves the best performance on various testsets at different CS ratios compared with the existing DL based method. The proposed CSformer is the first work to extend vision transformer to CS, and it has shown great potential to improve the CS performance.


%

\appendices
 
\ifCLASSOPTIONcaptionsoff
  \newpage
\fi



\bibliographystyle{IEEEtran}
\bibliography{IEEEabrv,bib_TranGp}
%



%




\end{document}